\documentclass[letter]{article}
\usepackage{graphicx} % Required for inserting images
\usepackage{authblk}
\usepackage{geometry}
\usepackage[square, numbers]{natbib}
\usepackage{amsmath,amsfonts,physics,relsize,slashed,amssymb}
\usepackage{graphicx,xcolor}
\usepackage[hidelinks]{hyperref}
\usepackage[capitalize]{cleveref}
\usepackage{comment}
\def\iu{\mathrm{i}}

\def\bar{\overline}

\title{Collective excitations and low-energy ionization signatures of relativistic particles in silicon detectors} 

\author[1]{Rouven Essig}
\author[2]{Ryan Plestid}
\author[1,3,$\dagger$]{Aman Singal}

\affil[1]{C. N. Yang Institute for Theoretical Physics, Stony Brook University, Stony Brook, NY 11794, USA}
\affil[2]{Walter Burke Institute for Theoretical Physics, California Institute of Technology, Pasadena, CA 91125, USA}
\affil[3]{Institute for Advanced Computational Sciences, Stony Brook University, Stony Brook, NY 11794, USA}
\affil[$\dagger$]{\href{mailto:aman.singal@stonybrook.edu}{aman.singal@stonybrook.edu}}
\date{}

\begin{document}
\begin{flushright}
    CALT-TH-2024-03, YITP-SB-2023-40
\end{flushright}

\begin{minipage}[h]{\textwidth}
    \maketitle    
\end{minipage}

\begin{abstract}
    Solid-state detectors with a low energy threshold have several applications, including searches of non-relativistic halo dark-matter particles with sub-GeV masses. When searching for relativistic, beyond-the-Standard-Model particles with enhanced cross sections for small energy transfers, a small detector with a low energy threshold may have better sensitivity than a larger detector with a higher energy threshold. In this paper, we calculate the low-energy ionization spectrum from high-velocity particles scattering in a dielectric material.  We consider the full material response including the excitation of bulk plasmons. We generalize the energy-loss function to relativistic kinematics, and benchmark existing tools used for halo dark-matter scattering against electron energy-loss spectroscopy data. Compared to calculations commonly used in the literature, such as the Photo-Absorption-Ionization model or the free-electron model, including collective effects shifts the recoil ionization spectrum towards higher energies, typically peaking around 4--6 electron-hole pairs. We apply our results to the three benchmark examples: millicharged particles produced in a beam, neutrinos with a magnetic dipole moment produced in a reactor, and upscattered dark-matter particles. Our results show that the proper inclusion of collective effects typically enhances a detector's sensitivity to these particles, since detector backgrounds, such as dark counts, peak at lower energies. 
\end{abstract}

\vfill 
\pagebreak

\section{Introduction}

    Several new direct-detection concepts to search for halo dark matter particles with mass below the proton have been proposed over the past decade~\cite{Essig:2022dfa}. Since dark-matter particles in our Milky-Way halo are non-relativistic, with $\beta\equiv v/c\sim 10^{-3}$, the resulting events have very little energy. Fortunately, the theoretical progress has been accompanied by a new generation of ever-improving low-threshold solid-state detectors capable of sensing the low-energy signals. In particular, some of the leading direct-detection bounds on halo dark matter with sub-GeV masses are from experiments that use semiconductors, usually silicon, that search for dark matter particles interacting with electrons~\cite{Essig:2011nj}, e.g.,~\cite{DAMIC:2016lrs,Crisler:2018gci,Agnese:2018col,SENSEI:2019ibb,Aguilar-Arevalo:2019wdi,SENSEI:2020dpa,Amaral:2020ryn,Arnaud:2020svb,DAMIC-M:2023gxo}. Their main advantage over larger detectors and noble-liquid or molecular targets~\cite{Agnes:2018oej,Essig:2012yx,Essig:2017kqs,Angle:2011th,Aprile:2016wwo,XENON:2019gfn,XENON:2021qze,PandaX-II:2021nsg,Blanco:2019lrf} is their superior energy threshold, which translates into them having sensitivity to lower dark-matter masses. On the other hand, for semi-relativistic or relativistic dark-matter particles or other particles beyond the Standard Model (BSM), large-volume detectors often have an advantage over smaller solid-state detectors, as the low energy threshold of the small detectors does not compensate for the large volume of the bigger detectors. There are, however, several cases where a kinematic or dynamical enhancement occurs at low energies and for which a low-threshold solid-state detector could be superior (see, for example, current and projected bounds in Ref.~\cite{Emken:2024nox}). For this reason it is imperative to have accurate predictions for the low-energy signal of relativistic particles that scatter in solid-state detectors. The goal of this paper is to provide these accurate predictions. 

    The theoretical calculations for the scattering rates of \textit{non-relativistic} halo dark matter in solid-state materials have become increasingly accurate and precise~\cite{Essig:2011nj,Essig:2015cda,Hochberg:2021pkt,Knapen:2021run,Knapen:2021bwg,Griffin:2021exdm,Trickle:2022exdm,Dreyer:2023ovn,Peterson:2023jgy}, with Refs.~\cite{Knapen:2021run,Hochberg:2021pkt, Knapen:2021bwg} emphasizing the need to include the full material response, which is captured by the dielectric function or energy loss-function. Similarly, we here emphasize that accurate predictions for the low-energy signals from the scattering of \textit{relativistic} particle must include the full material response (see also~\cite{Lasenby:2021wsc}). For example, a highly boosted particle will produce large transverse electromagnetic fields, as compared to a non-relativistic probe which will interact dominantly through Coulomb modes. We highlight in particular how relativistic particles can excite bulk plasmons (quasiparticles describing the collective excitation of electrons in, e.g.,~semiconductors), and how this drastically impacts the shape of the expected recoil spectrum when compared to calculations that do not include the plasmon. 

    Plasmons have previously received attention in the context of dark matter direct detection as a detection channel in e.g.,~\cite{Kurinsky:2020dpb,Kozaczuk:2020uzb,Knapen:2020aky,Lasenby:2021wsc,Gelmini:2020xir,Boyd:2022tcn}, however primarily in the context of non-relativistic halo DM scattering or of tunable plasma haloscopes for axion and dark photon searches \cite{Lawson:2019brd, Gelmini:2020kcu}. However, plasmons are not dominantly excited in the scattering of non-relativistic halo dark matter. The situation changes qualitatively when one considers relativistic particles, in which case the plasmon is easily accessible, and low-energy events can be resonantly enhanced. This must be taken into account in order to predict the signal correctly and to derive accurate constraints. We note that Ref.~\cite{Kurinsky:2020dpb} considered quasi-relativistic scattering in the context of boosted millicharged dark matter such that the plasmon is kinematically accessible; however, non-relativistic formulae were used to calculate the scattering rates.

    In this paper, we focus on silicon as a representative material, but our results can be generalized to other dielectric materials including germanium, gallium arsenide, and silicon carbide. We consider in particular three types of particles and mediators: millicharged particles (produced in a beam), particles with a magnetic dipole moment (such as a neutrino with a magnetic dipole moment), and particles (including dark matter) interacting with a massive (heavy or ultralight) dark photon that is kinetically mixed with the ordinary photon. Each of these can be relativistic when interacting with a detector material: they can be produced in accelerator-based experiments or accelerated due to scattering in ``high-energy'' environments; examples for the latter include ``solar-reflected'' dark matter~\cite{An:2017ojc,Emken:2017hnp,Emken:2019hgy,Emken:2021lgc,An:2021qdl,Zhang:2023xwv,Emken:2024nox} and dark matter boosted by cosmic-ray scattering~\cite{Bringmann:2018cvk,Cappiello:2018hsu,Ema:2018bih,Dent:2019krz,Dent:2020syp}. 

\section{Methods}
    \subsection{Low-energy excitations from relativistic particle scatters} \label{Fermi-Review}
        \subsubsection{General formalism for energy loss of particles scattering with electrons}

            A general formalism for energy loss by electrically charged particles was first given in~\cite{Fermi1940}. The treatment assumes that the incident test charge, with four-momentum $p$, may be treated as a classical source of $\vb{E}$ and $\vb{B}$ fields. This is equivalent to the eikonal approximation~\cite{Weinberg_Textbook} (valid for $k\ll p$ with $k$ the four-momentum transfer to the sample). Small low-threshold detectors can only compete with large high-recoil detectors when their sensitivity is dominated by regions of low momentum transfer; therefore, we will be interested in energy transfers no larger than 50~eV and momentum transfers less than the Fermi momentum $k_F \sim 5~{\rm keV}$, such that the classical electrodynamics treatment is an extremely good approximation over the full kinematic regime of interest.

            We now give the scattering cross section for three different types of particles and mediators, before commenting on other cases. In all expressions, bulk material properties that differ from a free-electron cross section, i.e., ``collective effects'', are captured using the dielectric function $\epsilon(\omega,k)$. 

            \paragraph{1.~Particles with an electric charge.} 
            For an electrically charged particle, the resulting cross section (differential with respect to energy loss) is given by~\cite{Fermi1940,Allison:1980}
            \begin{equation}\label{energy-loss-fermi}
                \begin{split}
                    \dv{\sigma}{\omega} = &\frac{2\alpha\varepsilon^2}{n\pi \beta^2}\int_{k_{\rm min}}^{k_{\rm max}} \frac{\dd k}{k} ~\bigg\{\mathrm{Im}\left(-\frac{1}{\epsilon(\omega, k)}\right)+\left(\beta^2k^2 - \omega^2\right)\mathrm{Im}\left(\frac{1}{-k^2 + \epsilon(\omega, k)\omega^2}\right)\bigg\}~. 
                \end{split}
            \end{equation}
            Here $k=|\vb{k}|$ is the three-momentum transfer to the sample, $\varepsilon$ is the charge of the probe in units of $e$, $\omega$ is the energy transfer, $\beta=|\vb{p}|/E$ is the three-velocity of the probe particle in the rest frame of the detector, $n$ is the number density of atoms, and $\epsilon(\omega,k)$ is the dielectric function of the material. The minimum momentum transfer is set by $k_{\rm min} = \omega/\beta$, while the maximum momentum transfer is $k_{\rm max} = 2 |\vb{p}| - k_{\rm min}$; in practice, since the integral has negligible support for $k \gg k_F$, the upper limit of the integral, $k_{\rm max}$, can simply be taken to infinity. This is justified for $k_\mu k^\mu\ll 2 p_\mu k^\mu$, which is appropriate when $|\vb{p}|\gg k_F \sim 5~{\rm keV}$, since $\abs{\vb{k}} \lesssim k_F$ dominate the energy loss function. In Coulomb gauge, the two terms in the above equation may be identified with the exchange of Coulomb modes and transverse photons, respectively, and are related to density-density and current-current correlators in the rest frame of the material.

            \paragraph{2.~Particles with a magnetic dipole moment.} 
            The above formula can be generalized to other models beside a point-like electrically charged particle. The simplest generalization involves higher electromagnetic multipoles. For concreteness, we consider here a relativistic particle with a magnetic dipole moment $\mu$, which is described by $\mathcal{L}_{\rm int} \supset \frac{\mu}{2} \bar{\psi} \sigma_{\mu\nu} F^{\mu\nu} \psi $. In this case, since the mediator is still a Standard Model photon, one may obtain the correct energy loss formula from \cref{energy-loss-fermi} by comparing the lepton tensors for a magnetic dipole moment and a millicharge. In the limit of a highly-boosted incident particle, this ratio is simply $-\mu^2 k_\nu k^\nu/e^2= \mu^2(\vb{k}^2-\omega^2)/e^2$. Therefore, for the case of a neutrino magnetic moment, we find
            \begin{equation}\label{energy-loss-nu}
                    \dv{\sigma}{\omega} = \frac{1}{2n \pi^2 \beta^2}\int_{k_{\rm min}}^{k_{\rm max}} \frac{\dd k}{k} ~~\mu_{\nu_\alpha}^2(k^2-\omega^2)\times \bigg\{\mathrm{Im}\left(-\frac{1}{\epsilon(\omega, k)}\right)+\left(\beta^2 k^2 -\omega^2\right)\mathrm{Im}\left(\frac{1}{-k^2 + \epsilon(\omega, k)\omega^2}\right)\bigg\}\ ,
            \end{equation}
            where we have allowed the neutrino magnetic moment to depend on the neutrino flavor.

            \paragraph{3.~Particles interacting with a massive vector mediator (dark photon).} A similar procedure for a vector mediator with mass $m_V$ (i.e., a dark photon) with coupling $g_\chi$ to the relativistic probe with mass $m_\chi$, and coupling $g_e$ to electrons yields
            \begin{equation}\label{energy-loss-V}
                \begin{split}
                    \dv{\sigma}{\omega} &= \frac{2 \alpha}{n \pi \beta^2} \qty[\frac{g_eg_\chi}{4\pi \alpha} ]^2\int_{k_{\rm min}}^{k_{\rm max}} \frac{\dd k}{k} ~~\qty(\frac{k^2-\omega^2}{k^2+m_V^2-\omega^2})^2~\\
                    &\hspace{0.225\linewidth} \times \bigg\{\mathrm{Im}\left(-\frac{1}{\epsilon(\omega, k)}\right)+\left(\beta^2 k^2 -\omega^2\right)\mathrm{Im}\left(\frac{1}{-k^2 + \epsilon(\omega, k)\omega^2}\right)\bigg\}~,\\
                    &= \frac{1}{\beta^2}\overline{\sigma}_e \int_{k_{\rm min}}^{k_{\rm max}} \frac{\dd k}{k} ~~\qty(\frac{k^2-\omega^2}{k^2+m_V^2-\omega^2})^2~\frac{2 [(\alpha m_e)^2 + m_V^2 - \omega^2]^2}{(16\pi^2 \alpha)\mu_{\chi e}^2 n } \\
                    &\hspace{0.225\linewidth} \times \bigg\{\mathrm{Im}\left(-\frac{1}{\epsilon(\omega, k)}\right)+\left(\beta^2 k^2 -\omega^2\right)\mathrm{Im}\left(\frac{1}{-k^2 + \epsilon(\omega, k)\omega^2}\right)\bigg\}~,
                \end{split}
            \end{equation}
            where $\mu_{\chi e} = m_e m_\chi / (m_e + m_\chi)$ is the reduced mass. Strictly speaking we assume a vector coupling proportionally to the electric charge, i.e., such that the coupling to protons is $g_p=-g_e$. For practical purposes, if the interaction couples to electrons at all (which dominate material responses), the formulae can often still be applied, unless the coupling to nucleons is much larger than the coupling to electrons. In the second equation, we have re-written the cross section in terms of a ``reference cross section'', $\overline{\sigma}_e$, as is common in the dark-matter literature. A comparison between Eq. \ref{energy-loss-fermi} and Eq. \ref{energy-loss-V} shows that the case of particles interacting with a vector mediator corresponds to the case of millicharged particles in the limit $m_V \rightarrow 0$ and $\displaystyle \abs{\varepsilon} = \abs{\frac{g_eg_\chi}{4\pi \alpha}}$.

            \paragraph{4.~Other particles and interactions.} More generally, one may consider different types of mediator particles, for instance a massive scalar, vector, pseudoscalar, or pseudovector. In this case one should be careful to treat in-medium effects properly, which can be accomplished using a thermal field theory formalism as discussed in Ref.~\cite{Catena:2024rym} and in Appendix A of~\cite{Lasenby:2021wsc}. Since the response of metals and semi-conductors are dominated by valence electrons, one may to a good approximation consider the charge and electron density as interchangeable and argue on these grounds that electromagnetic data usefully constrains any model of electrophilic interactions. This is especially true in the limit of small momentum transfers, where a non-relativistic approximation can be employed for the electrons and protons that dictate the detector response. In this limit, the longitudinal component of the electromagnetic response function is related to the scalar response function~\cite{Lasenby:2021wsc}, and the same energy loss function characterizes both scalar- and vector-mediated scattering. This approximation is valid even for relativistic probes provided the momentum transfer satisfies $\omega, k\ll m_e$. These constraints are satisfied for all of the phenomenology we consider here, and so our results apply to both light scalar and vector mediators. \\
            Pseudoscalar and pseudovector interactions lead to spin-density-dependent response functions at low momentum transfers that cannot be extracted using EELS data. These could be obtained empirically using neutron magnetic scattering~\cite{Neutron_Textbook}, however we do not pursue this idea further here. We focus instead on models whose required detector response can be obtained from standard EELS measurements. This is well motivated since light vector and scalar mediators naturally give cross sections that are enhanced in the low-$q^2$ limit where low-threshold detectors are most effective. 

            With the formulae in hand for the scattering cross section in semiconductors for various particles, \cref{energy-loss-fermi,energy-loss-nu,energy-loss-V}, we see that the problem reduces to finding accurate expressions or data for the dielectric function, $\epsilon(\omega,k)$. We discuss theoretical and experimental estimates for the dielectric function in the subsection titled `expressions for the dielectric function and comparison with EELS data.' We will see that the plasmon peak plays a crucial role in determining the differential spectrum. Before doing so, however, we comment in the subsection titled `comparison with average energy loss formalism' on how our formulae compare with the ``average energy loss'' formalism commonly used in the literature. \Cref{fig:k_W} of supplementary materials shows the structure of $k^n \mathrm{Im}\left\{-1/\epsilon(\omega, k)\right\}$ for different values of $n$. \Cref{fig:om_W} of supplementary materials shows the appearance of plasmon for
            $$I_n(\omega) = \int \dd k~k^n~\mathrm{Im}\left\{\frac{-1}{\epsilon(\omega, k)}\right\}, $$
            by showing results for $n = -1, 0, \dots, 4$. It is apparent that the plasmon is more relevant for longer range (lower $k$) interactions. 

        \subsubsection{Comparison with average energy loss formalism} \label{subsec:average-energy-loss}

            \Cref{energy-loss-fermi} forms the basis of the theory of average energy loss for ultra-relativistic particles~\cite{PDG}. Indeed, weighting $\dd \sigma /\dd \omega$ by the energy transfer $\omega$, and integrating over available energy losses one can derive expressions for $\langle \dd E/\dd x\rangle$. In many contexts involving Standard Model particles and relatively thick targets, the average energy loss is the relevant quantity. Exceptions to this rule exist even within the Standard Model. For example in the thin-target limit, it is well known that the most-probable, as opposed to the mean, energy loss is a better characterization of energy loss~\cite{PDG,Landau,Landau-Vavilov}. More precisely, energy loss is probabilistic and characterized by the Vavilov distribution~\cite{Landau-Vavilov}. The inequivalence of these two quantities stems from the fact that Mott scattering is governed by a power-law with a long tail such that $\langle \dd E/\dd x\rangle$ receives $O(1)$ contributions from high-energy scatters that will rarely, if ever, occur for a fixed number of scatters against a thin target. As a result, the energy loss distribution in thin targets is better characterized in terms of its mode (i.e., its most likely energy loss) as opposed to its mean~\cite{PDG}.

            An analogous issue appears when one considers the detection of feebly interacting particles. For example, the cross section $\dd \sigma/\dd \omega$ for a millicharged particle is obtained by re-scaling the Standard Model cross section by $\varepsilon^2$, and so naively the statistical properties of the two distributions are identical. In practice this is not the case, because the microscopic cross section for energy loss is a fat-tailed distribution. This makes the average energy loss, which determines the mean of the Gaussian distribution that emerges by the central limit theorem and characterizes energy loss for a particle with charge $e$, a poor characterization of the distribution that controls millicharged particle detection. 

            When considering the detection of feebly interacting particles, it is therefore essential to properly model the scattering cross section as a function of energy loss, and in particular the location of its peak. Approximations that model well the average energy loss $\langle \dd E/\dd x\rangle$, such as the photo-absorption ionization (PAI) model \cite{Bichsel:2006cs,TEXONO:2018nir} as originally defined in \cite{Allison:1980}, are a poor choice for studies of detector sensitivity to feebly interacting particles. The PAI model makes crude assumptions that completely mismodel $\dd\sigma/\dd \omega$, and predicts a peak in the distribution at $\sim 5~{\rm eV}$ rather than the correct value of $\sim 16~{\rm eV}$ i.e., at the plasmon peak (see the `Millicharged particles' subsection in the `Results' section). The ultimate sensitivity of a detector to feebly interacting particles is governed by the locations in phase space in which the detection cross section is maximal. In a conventional large-volume detector, such as a liquid scintillator or a noble gas detector, which have a relatively high energy threshold (above the peak of $\dd\sigma/\dd \omega$), the highest event rate will always occur at the lowest possible recoil energies (see e.g.~\cite{Magill:2018tbb,Harnik:2019zee,Harnik:2020ugb} for a discussion). For low-threshold detectors, with sensitivity to energies at the plasmon peak and below, such as Skipper-CCDs~\cite{Tiffenberg:2017aac}, the greatest sensitivity is obtained close to the plasmon peak when it is kinematically accessible. This dictates the expected event spectrum and how to optimize the cuts for experimental searches. We will see several examples in the `Results' section. 

    \subsection{Expressions for the dielectric function and comparison with EELS data \label{Tests}}

        We saw in the subsection titled `low-energy excitations from relativistic particle scatters,' that the differential scattering cross sections can be expressed in terms of the dielectric function of the material; in particular, \cref{energy-loss-fermi} gives the expression for an electrically charged particle, \cref{energy-loss-nu} is applicable for particles interacting with a magnetic dipole moment, and \cref{energy-loss-V} for particles with a dark-photon mediator. The dielectric function captures all relevant collective effects, and the formulae \cref{energy-loss-fermi,energy-loss-nu,energy-loss-V} are valid for both relativistic and non-relativistic kinematics. We here discuss theoretical approximations for the dielectric function, and will show that we can use EELS data to validate our expressions, at least for electrically charged particles. 

        There are several publicly available tools for calculating dark-matter scattering off various materials and the resulting direct-detection signals, including QEDark~\cite{Essig:2015cda,QEDark}, DarkELF~\cite{Knapen:2021run,Knapen:2021bwg}, EXCEED-DM~\cite{Griffin:2021exdm,Trickle:2022exdm}, and QCDark~\cite{Dreyer:2023ovn,QCDark}. Due to the kinematics of virialized dark matter, the focus of the community has been to characterize correctly the region of phase space in which $k\gtrsim \beta_{\rm vir}^{-1} \omega$ where $\beta_{\rm vir}\sim 10^{-3}$, such that the typical values for the recoil energy and momentum transfer are $\omega \sim {(\rm few)~ eV}$ and $k\sim {\rm (few)~keV}$, respectively.

        The situation differs substantially for relativistic particles with $\beta \sim 1$. Then the natural scaling is $\omega \sim k \sim {(\rm few)~ eV}$. Often one encounters discussions of the so called ``optical limit,'' which refers to $k \ll k_F$, or tacitly $k\rightarrow 0$~\cite{Bardasis:1967eps, Hybertsen:1987eps}. Indeed, if one considers optical absorption measurements of on-shell photons then $\omega = k\sim {\rm few~eV}$. There is a crucial difference, however, between optical absorption and the scattering of relativistic particles. Photons are always transversely polarized, whereas a charged particle can interact with the sample via longitudinal Coulomb modes. In fact, it is precisely these longitudinal modes, i.e., the longitudinally polarized bulk plasmon, which dominate the response function. Optical absorption data is therefore a \textit{poor proxy} for relativistic scattering of charged particles~\cite{Kundman_thesis}. 

        We can write the dielectric function in terms of its real and imaginary part, 
        \begin{equation}
            \epsilon(\omega,k) =  \epsilon_1(\omega,k) + i \epsilon_2(\omega,k)\,,
        \end{equation}
        where $\epsilon_{1,2}(\omega,k)$ are real, so that the energy loss function can be written as
        \begin{equation}
            {\rm Im}\left(\frac{-1}{\epsilon_1(\omega,k) + \iu \epsilon_2(\omega,k)}\right) = \frac{\epsilon_2(\omega,k)}{|\epsilon_1(\omega,k)|^2 + |\epsilon_2(\omega,k)|^2}\,.
        \end{equation}

        \begin{figure}
            \centering
            \includegraphics[width=\linewidth]{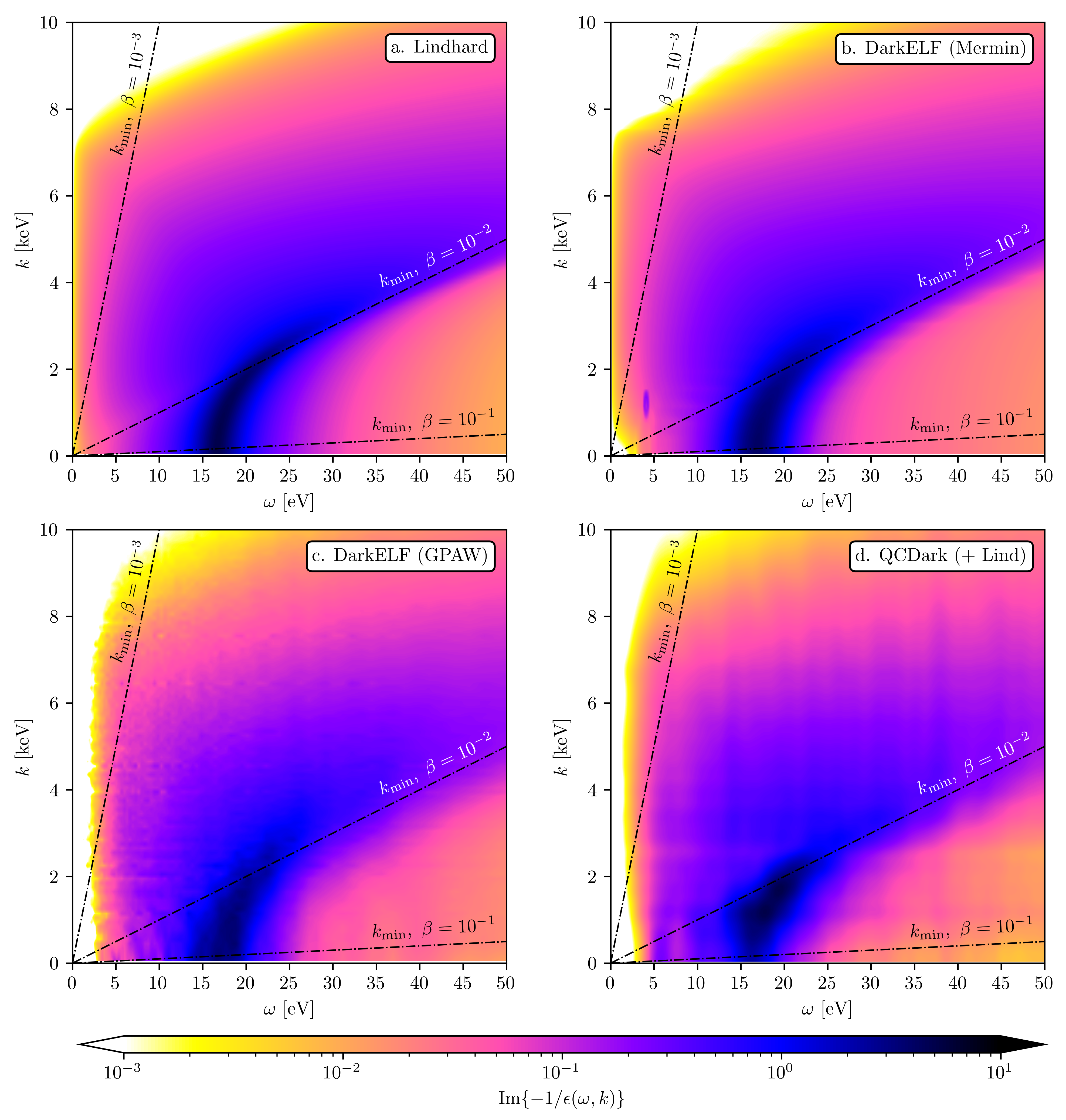}
            \caption{\textbf{Silicon electron loss function approximations.} The electron loss function, $\mathrm{Im}\left\{-1/\epsilon(\omega, k)\right\}$, for silicon are plotted as a function of $\omega$ and $k$. Panel a shows the Lindhard approximation to the electron loss function; panel b shows the Mermin approximation built into the \texttt{DarkELF} program~\cite{Knapen:2021bwg, Knapen:2021run}; panel c shows the GPAW DFT calculation of the dielectric function built into \texttt{DarkELF}; panel d shows the \texttt{QCDark} calculation of the imaginary part of the dielectric function~\cite{Dreyer:2023ovn} screened by the Lindhard $1/\abs{\epsilon(\omega, k)}^2$ as given by \cref{ansatz}. Panel d, due to the low $\mathbf{k}-$grid used in the calculation, is Gaussian smoothed with $\sigma_\omega = 0.5~{\rm eV}$ and $\sigma_k = 200~{\rm eV}$. In all plots, the plasmon is visible for low $k$ and $\omega \sim 16.6~{\rm eV}$. The lines indicate the minimum momentum, $k_\mathrm{min}$, required to transfer energy $\omega$ from an incoming particle with speed $\beta$. Note that the plasmon threshold occurs close to $\beta \sim 10^{-2}$.}
            \label{fig:eps}
        \end{figure}

        Many existing tools in the literature for dark matter scattering do not properly model the real part of the dielectric function at small momentum transfers. Fortunately, a good qualitative and reasonable quantitative description of $\epsilon(\omega,k)$ for $k\ll k_F$ is given by the Lindhard model~\cite{Lindhard}, which provides a reasonable description for millicharged particle searches that are dominated by low momentum transfers. For massive mediators, or higher dimensional operators such as a magnetic dipole moment, larger momentum transfers can contribute $O(1)$ fractions to the total cross section. Fortunately, in the limit $k \gtrsim k_F$ where the Lindhard model is unreliable~\cite{Cohen_Louie_2016}, the absolute value of the dielectric function is approximately unity (since $\epsilon_1\approx 1$ and $\epsilon_2 \approx 0$) which is properly reproduced by the Lindhard model. The Lindhard model does not, however, properly model $\epsilon_2$ at large values of $k$. Hence the Lindhard model is able to produce a reliable value for $\abs{\epsilon(\omega, k)}^2$ while being unable to calculate the electron loss function $\left\{-1/\epsilon(\omega, k)\right\}$ at large $k\gtrsim k_F$. 

        More robust methods of calculating the dielectric function involve calculation of the electronic wavefunction using density functional theory (DFT)~\cite{Kohn:1965DFT, Hohenberg:1964DFT, Cohen_Louie_2016}, and employing the random phase approximation (RPA)~\cite{Bohm:1951RPA1, Bohm:1952RPA2, Bohm:1953RPA3, Ehrenreich:1959RPA4}. The open source tool \texttt{DarkELF}~\cite{Knapen:2021bwg, Knapen:2021run} includes RPA calculations of both the real and imaginary parts of the dielectric function calculated using the DFT software \texttt{GPAW}. The \texttt{DarkELF} package also includes the option to fit a superposition of Lindhard dielectrics, which is referred to as the `Mermin' model. The current implementation \texttt{DarkELF} does not, however, reconstruct the wavefunctions of core electrons. This causes the valence electron wavefunctions from \texttt{DarkELF} to mismodel large-$k$ modes. A reliable description of large-$k$ modes can be obtained using other tools, for example \texttt{QCDark}~\cite{Dreyer:2023ovn}. In its current state, however, \texttt{QCDark} only calculates the imaginary part of the RPA dielectric function. Therefore, when using \texttt{QCDark} to predict the energy loss function, one must supply the real part of the dielectric externally. More specifically, \texttt{QCDark} provides the crystal form factor as developed in~\cite{Essig:2015cda}, which is equivalently written as
        \begin{equation}
            \abs{f_\mathrm{crystal}(k, \omega)}^2 = \frac{k^5 V_\mathrm{cell}}{8\pi^2\alpha_{EM}^2m_e^2} \epsilon^\mathrm{RPA}_2(\omega, k)\,,
        \end{equation}
        where $V_\mathrm{cell}$ is the volume of the unit cell, $\alpha_{EM}$ is the electromagnetic fine structure constant, and $m_e$ is the mass of the electron. We find that a reliable {\it global} approximation of the energy loss function can be obtained for silicon by taking the following model 
        \begin{equation}\label{ansatz}
            {\rm Im}\bigg[\frac{-1}{\epsilon_1(\omega,k) + \iu \epsilon_2(\omega,k)}\bigg]_{\rm model} = \frac{\big[\epsilon_2(\omega,k)\big]_{\rm QCDark} }{\big[\qty|\epsilon(\omega,k)|^2\big]_{\rm Lindhard}}~. 
        \end{equation}
        This model works in the low-energy region near the plasmon peak because the Lindhard model is reliable there and agrees reasonably with the more sophisticated calculation of $\epsilon_2$ from \texttt{QCDark}~\cite{Dreyer:2023ovn}. In the high-energy region, the imaginary part of the dielectric function is small, $\epsilon_2 \ll \epsilon_1 \approx 1$, such that the denominator can effectively be replaced by unity in any model. We are then entirely insensitive to the mismodeling of $\epsilon_2(\omega,k)$ at large values of $k$ in the Lindhard model, while simultaneously benefiting from the Lindhard model's realistic description of the bulk plasmon. However, note that the low $k$ electron loss function is underestimated using this method, likely due to an overestimate of $\abs{\epsilon(\omega, k \ll k_F)}^2$ from the Lindhard model.

        In \cref{fig:eps}, we show the electron loss function, which dominates energy loss in both the non-relativistic and relativistic regimes, computed in various approximations. This includes the Lindhard model~\cite{Lindhard}, the Mermin model and density functional theory calculations from \texttt{DarkELF}~\cite{Knapen:2021run, Knapen:2021bwg}, and \texttt{QCDark}~\cite{Dreyer:2023ovn} screened by the Lindhard dielectric function as given in \cref{ansatz}.

        \begin{figure}[t]
            \centering
            \includegraphics[width=0.495\linewidth]{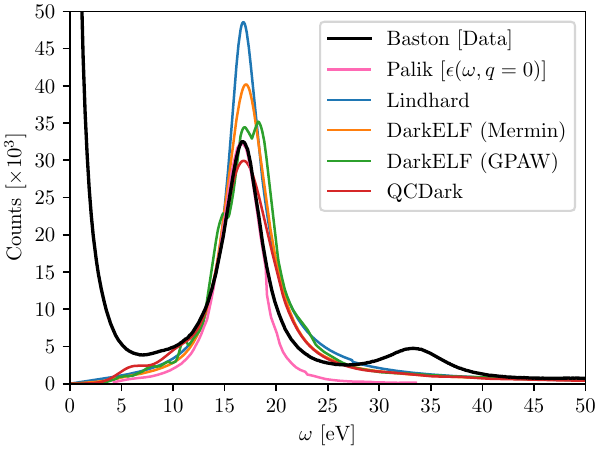}
            \caption{\textbf{Comparison with Electron Energy-Loss Spectrocopy (EELS).} Data (\textbf{black line}) are compared to our theoretical modeling (various \textbf{colored lines}) in a bulk silicon from~\cite{Baston:EELS, Ewels:2016EELSDB} with an incident electron beam kinetic energy of $T = 100~$keV. We model the theoretical EELS rates by calculating the EELS cross section using \cref{energy-loss-fermi} with various different approximations for the dielectric function $\epsilon(\omega, k)$. We normalize the rates to the plasmon peak, $\omega = \omega_{pl}$ using experimental $\epsilon(\omega, k \rightarrow 0)$ from~\cite{Palik_book}, in conjunction with \cref{eq:long_wavelength_EELS} and \cref{eq:flux_fix}, plotted as \textit{pink} line. The EELS data includes multiple scattering, with a secondary scattering peak visible at $\omega = 2\omega_{pl}$, which can be included by modeling multiple scatters (see~\cite{Kundman_thesis} for a discussion). Note that the sharply rising peak in the experimental data towards low energies, $\omega \lesssim 10~$eV is due to elastic scatters between incident electrons and the lattice, and is not captured in the models of the dielectric function $\epsilon(\omega, k)$ considered here. \textit{Blue} line corresponds to the Lindhard model~\cite{Lindhard}, \textit{orange} and \textit{green} lines corresponds to \texttt{DarkELF} (Mermin) and \texttt{DarkELF} (GPAW) respectively \cite{Knapen:2021bwg, Knapen:2021run}, and the red line corresponds to \texttt{QCDark} \cite{Dreyer:2023ovn}.}
            \label{fig:eels}
        \end{figure}

        The response functions we make use of here are easily compared with existing EELS data. An EELS experiment involves quasi-relativistic electrons impinging on a thin sample (which precisely replicates the kinematics we are interested in) and allows for both Coulomb-mode and transverse-mode mediated scattering~\cite{Kundman_thesis}. Any tool claiming to reliably predict low-energy event rates in a silicon (or other semiconductor) detector must necessarily reproduce EELS spectra. Previous comparisons with EELS (for URu$_2$Si$_2$ specifically) and X-ray scattering data have focused on the kinematic regime relevant for halo dark matter direct detection i.e., $k\sim 5~{\rm keV}$~\cite{Hochberg:2021pkt}.

        \Cref{fig:eels} shows the EELS experimental data from Refs.~\cite{Baston:EELS, Ewels:2016EELSDB} for silicon. The incident electrons have a kinetic energy of $100~{\rm keV}$. The peak at $\omega \sim 0$ corresponds to elastic scattering of incident electrons with the lattice, and hence does not correspond to electron-hole pair excitations. The plasmon peak at $\omega = \omega_{pl}\approx 16.6~{\rm eV}$ is visible. We further use experimental data available for the dielectric function at long wavelengths, $\epsilon(\omega) \equiv \epsilon(\omega, k\rightarrow 0)$ from~\cite{Palik_book} to compute the differential EELS cross section~\cite{Egerton2011}, 
        \begin{equation}\label{eq:long_wavelength_EELS}
            \dv{\sigma}{\omega} \approx \frac{\alpha}{n\pi \beta^2}\mathrm{Im}\left\{\frac{-1}{\epsilon(\omega)}\right\}\log\left\{1 + \frac{\theta^2}{\theta_E^2}\right\}.
        \end{equation}
        Here $\theta$ is the collection angle for the data (1.6 mrad) and $\theta_E = \omega/\gamma m_e\beta^2$. We then normalize the counts $C(\omega)$ observed as
        \begin{equation}\label{eq:flux_fix}
            C(\omega) = \kappa \dv{\sigma}{\omega}
        \end{equation}
        to fix the electron flux, where we obtain $\kappa$ by fixing the value of $C(\omega_{\rm pl})$ to match experimental data at the plasmon peak.

        We compare the data to the cross section computed using the Lindhard model~\cite{Lindhard}, Mermin and GPAW calculations from \texttt{DarkELF}~\cite{Knapen:2021run, Knapen:2021bwg}, and \texttt{QCDark}~\cite{Dreyer:2023ovn} screened with Lindhard as in \cref{ansatz}. We normalize the counts using the flux obtained by fitting the semi-empirical differential cross section using \cref{eq:long_wavelength_EELS} and \cref{eq:flux_fix}. Any EELS measurement will have multiple scattering peaks whose amplitude grow with increased sample thickness~\cite{Kundman_thesis}. In this work, we ignore multiple scatterings, though they can be included by assuming a Poisson process and normalizing the second peak to the corresponding peak in the experimental data. This would then fix the thickness of the sample.

        The Lindhard and Mermin models~\cite{Lindhard, Knapen:2021bwg, Knapen:2021run} overestimate the cross section, while \texttt{QCDark} (with Lindhard screening \cref{ansatz})~\cite{Dreyer:2023ovn} slightly underestimates it. %The latter is likely due to the Lindhard model overestimating the screening, causing the electron loss function to be underestimated at $k \ll k_F$. 
        The \texttt{DarkELF} (GPAW) dielectric function~\cite{Knapen:2021bwg, Knapen:2021run} only slightly overpredicts the peak of the EELS data, which means the differential cross section agrees with the semi-empirical cross section that uses~\cite{Palik_book} and \cref{eq:long_wavelength_EELS}. However, it predicts a split plasmon with peaks at $\omega \sim 16.9 ~{\rm eV}$ and $\omega\sim 18.2 ~ {\rm eV}$, which seems unphysical, and a slightly broader plasmon peak than other approximations. 

\section{Results \label{Pheno}}

    In what follows we focus on three representative examples of BSM models for which low-threshold silicon detectors are well suited. We focus on the modification of the cross section due to collective effects. First, we consider millicharged particles produced in accelerator beams and/or cosmic rays~\cite{Magill:2018tbb,Harnik:2019zee,ArgoNeuT:2019ckq,Plestid:2020kdm,Oscura:2023qch}. Millicharged particles have a cross section that is enhanced at small momentum transfers, and, in the contexts we consider, are highly boosted. They are therefore a prime example where it is important to include plasmon excitations to correctly model the expected recoil spectrum. Moreover, searches for millicharged particles in the 100 MeV--100 GeV range is an active area of research, and SENSEI has recently demonstrated excellent sensitivity from data taken in the MINOS hall at Fermilab~\cite{SENSEI:2023gie}; moreover, Oscura will have sensitivity to such particles~\cite{Oscura:2023qch}. Second, we consider a small silicon detector near a nuclear reactor as has recently been proposed, for example, in the context of the vIOLETA collaboration~\cite{Fernandez-Moroni:2020yyl, Fernandez-Moroni:2021nap, CHANDLER:2022gvg}. As a representative example, we consider a search for a neutrino magnetic dipole moment, where scattering is moderately enhanced at low momenta, but not as strongly as for millicharged particles. The same set-up has promising sensitivity to light mediators that couple neutrinos to electrons and nucleons. Finally, we consider boosted dark matter as may be produced via solar reflection or by cosmic ray upscattering, e.g.~\cite{Emken:2024nox,An:2017ojc,Emken:2017hnp,Emken:2019hgy,Emken:2021lgc,An:2021qdl,Zhang:2023xwv,Bringmann:2018cvk,Cappiello:2018hsu,Ema:2018bih,Dent:2019krz,Dent:2020syp}. We study, in particular, how the sensitivity changes as the mass of the mediator is varied. 

        \begin{figure}
            \centering
            \includegraphics[width=\linewidth]{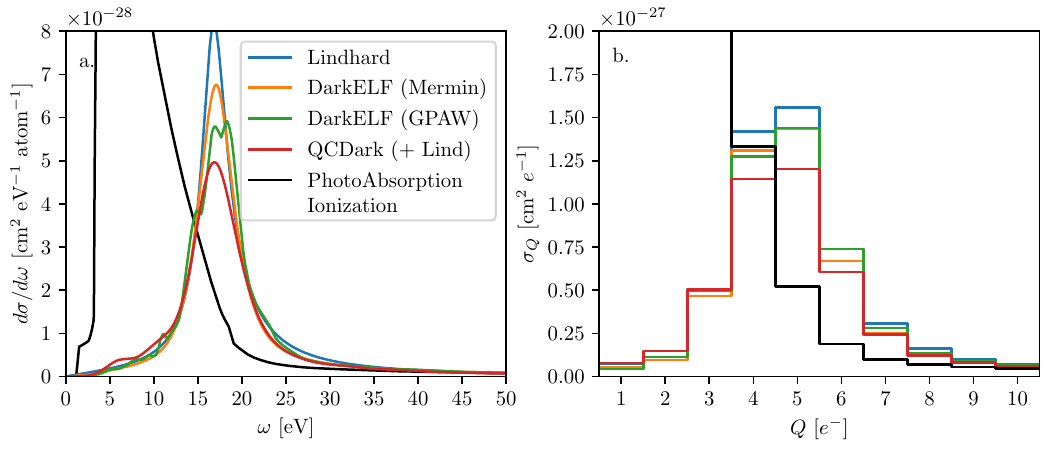}
            \caption{\textbf{Millicharged particles interacting with electrons, $\varepsilon = 10^{-4}$.} Panel a: the differential cross section of millicharged particle with charge $\varepsilon = 10^{-4}\ e^-$ scattering in a silicon target, given in units of cm$^2$ eV$^{-1}$ per unit cell. \textit{Blue} line corresponds to the Lindhard model~\cite{Lindhard}, \textit{orange} and \textit{green} lines corresponds to \texttt{DarkELF} (Mermin) and \texttt{DarkELF} (GPAW) respectively \cite{Knapen:2021bwg, Knapen:2021run}, and the red line corresponds to \texttt{QCDark} \cite{Dreyer:2023ovn}. The black line assumes the Photo Absorption Ionization (PAI) model~\cite{Bichsel:2006cs,TEXONO:2018nir} in which the dielectric function is inferred exclusively from optical measurements. The \texttt{QCDark} rates are calculated by screening the \texttt{QCDark} Im$\left\{\epsilon(\omega, k)\right\}$ with a Lindhard $\abs{\epsilon(\omega, k)}^2$ (see \cref{ansatz}), and are Gaussian smoothed with $\sigma_\omega = 0.5~{\rm eV}$, while the \texttt{DarkELF} and Lindhard lines use a fully self-consistent $\epsilon(\omega,k)$. Panel b: the cross section calculated as a function of the number of electron-hole pairs created, where we use the secondary ionization model from~\cite{ramanathan_ionization_2020} at 100~K. Note the peaks at $Q = 4~e^-$ and $Q = 5~e^-$.}
            \label{fig:mcps}
        \end{figure}

    \subsection{Millicharged particles}\label{subsec:millicharged-pheno}
        
        Accelerator-based production of low-mass millicharged particles leads to a flux that is almost entirely relativistic, assuming the incoming proton-beam energy is sufficiently large~\cite{Magill:2018tbb,Harnik:2019zee}. The cross section in this limit is nearly independent of the precise boost of the millicharged particle, $\gamma_{\rm mcp}$ and one may approximate the rate of millicharged particles that scatter downstream by 
        \begin{equation}
            \frac{\dd \Gamma_{\rm det}}{\dd \omega} = \Phi \times \frac{\dd\sigma_{\rm mcp} }{\dd \omega}~, 
        \end{equation}
        with $\dd \sigma_{\rm mcp}$ given by \cref{energy-loss-fermi}, and where $\Phi$ is the flux of relativistic millicharged particles. The integration measure $\dd k/k = \dd \log k$ is scale-independent such that small-$k$ regions are not phase-space suppressed. In $\log(k)$-space, the plasmon appears for $k\lesssim 3~{\rm keV}$ (as can be seen in \cref{fig:eps}). Since $k_{\rm min} \geq \omega$ for $\omega \sim 10~{\rm eV}$, the plasmon contributes appreciably to the integral for roughly two-decades in $k$-space. 

        Panel a of \cref{fig:mcps} shows the differential cross section per atom for a millicharged particle ($\varepsilon = 10^{-4}$) interacting with a silicon detector for various models (in 1~kg of silicon, there are $\sim 2.14\times10^{25}$ atoms). The black line shows the results for the Photo Absorption Ionization (PAI) model, which is frequently used to model energy loss of fast charged particles in gases and other materials~\cite{Allison:1980,SMIRNOV2005474}. Panel b of \cref{fig:mcps} shows the cross section as a function of the number of electron-hole pairs being created, $Q$, with the secondary ionization model taken from~\cite{ramanathan_ionization_2020}. We find that, compared to the PAI model, the corrected cross-section peaks at the plasmon peak, $\omega_p$, which leads to better detector sensitivity at higher charge ionizations, $Q\sim 5~e^-.$

        Our results can be immediately applied to set bounds on millicharged particles using existing data and to make sensitivity projections for future data. In particular, our cross sections were used by SENSEI to search for millicharged particles produced in the NuMI beam at Fermilab~\cite{SENSEI:2023gie}. The search was consistent with a null signal, and the constraints on millicharged particles was found to be world leading for certain millicharged-particle masses. The SENSEI analysis was based on data taken in 2020, which had previously been used to constrain sub-GeV dark matter interactions in~\cite{SENSEI:2020dpa}. The analysis in 2020 only included the bins containing $Q = 1-4~e^-$. Given that the events peak at $Q = 4~e^-$ and $5~e^-$, and also contain an appreciable number of events with $Q = 6~e^-$, SENSEI added the $Q = 5~e^-$ and $Q=6~e^-$ bins.  In addition, larger detectors are being planned. In particular, there is a plan to place a 1~kg Skipper-CCD detector in the NUMI beam line as part of the Oscura Integration Test before constructing the 10~kg Oscura dark matter detector. The cross sections discussed in this section are again needed for deriving accurate sensitivity projections~\cite{Oscura:2023qch}. 
    
    \subsection{Neutrino dipole moments}

        \begin{figure}
            \centering
            \includegraphics[width=\linewidth]{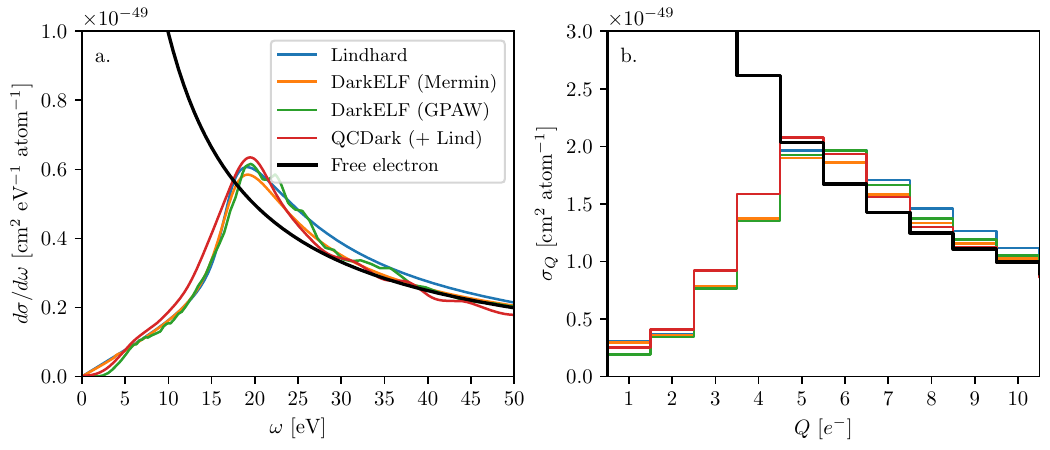}
            \caption{\textbf{Neutrinos interacting with electrons via magnetic dipole moment.} Panel a: the differential cross section of neutrinos interacting with a silicon target through a magnetic dipole moment $\mu_\nu = 10^{-12}~\mu_B$. \textit{Blue} line corresponds to the Lindhard model~\cite{Lindhard}, \textit{orange} and \textit{green} lines corresponds to \texttt{DarkELF} (Mermin) and \texttt{DarkELF} (GPAW) respectively \cite{Knapen:2021bwg, Knapen:2021run}, and the red line corresponds to \texttt{QCDark} \cite{Dreyer:2023ovn}. The black line shows the results for a free electron. Note that the free electron approximation is a good estimate for the shape of the differential cross-section compared to modeling the same via a dielectric approximation at high recoil energies, $\omega > 20$~eV, however it overestimates the rate by $\sim 50\%$ in this region. Panel b: the cross section versus the number of electron-hole pairs created, where we use the secondary ionization model from~\cite{ramanathan_ionization_2020} at 100~K. The charge ionization spectrum is drastically different between the correct calculation and the free-electron calculation, and $\sigma_Q$ first rises as a function of $Q$, until about $Q\sim 5~e^-$, then decreases again. }
            \label{fig:neutrinos}
        \end{figure}
    
        The magnetic dipole moments of the three neutrino species are predicted to either vanish, or be unobservably small, in the Standard Model~\cite{Fujikawa:1980yx,Giunti:2008ve,Giunti:2014ixa}. Searches for a non-zero neutrino dipole moment then represent a low-energy test of the Standard Model and, by proxy, an avenue for the discovery of BSM physics. Since a dipole operator is dimension-5, as compared to the standard dimension-6 contact interaction that governs neutrino scattering below the weak scale, it may be fruitfully pursued at low energies~\cite{Giunti:2014ixa}. 

        The recently proposed reactor neutrino experiment vIOLETA~\cite{Fernandez-Moroni:2020yyl,Fernandez-Moroni:2021nap}, aims to place a low-threshold Skipper-CCD near an operating nuclear reactor. One proposed use-case for vIOLETA is to search for anomalous signatures of a neutrino dipole moment and of light mediators that allow neutrinos to interact with other Standard Model particles, such as electrons. Viable signatures include coherent scattering on nuclei and scattering on electrons. 
        For dipole interactions and light mediators, the cross sections are comparable, and electron scattering is an attractive detection signature. Our results in \cref{energy-loss-nu} can be immediately applied to this detection channel.

        \Cref{fig:neutrinos} shows the differential cross section of neutrinos interacting with a silicon target for a magnetic dipole moment of $\mu_\nu = 10^{-12}~\mu_B$. Using \cref{energy-loss-nu} as opposed to assuming scattering off a free electron (used in~\cite{Fernandez-Moroni:2021nap}). 
        Collective effects alter both the overall rate, and the shape of the differential distribution with respect to energy transfer. %The shift in the overall normalization has straightforward implications for detection prospects (they are slightly weakened), however the 
        The altered energy transfer spectrum has non-trivial effects; shape discrimination is a powerful tool for distinguishing signal from background~\cite{Fernandez-Moroni:2021nap}, and the notable peaked structure visible in panel a of \cref{fig:neutrinos} offers a distinctive feature that may aid in future searches for neutrino dipole moments (or light mediators). This is to be compared with the free electron (at rest) recoil spectrum (assuming four valence electrons per atom), 
        \begin{equation}
            \qty[\dv{\sigma}{E_e}]_{\rm free~electron}= \alpha \mu_{\nu}^2\qty[\frac{1}{E_e}-\frac{1}{E_\nu}]~,
        \end{equation}
        which has no such feature and monotonically increases as $E_e \rightarrow 0$. 

        Panel b of \cref{fig:neutrinos} shows the cross section per atom of silicon as a function of electron-hole pairs ionized, using the secondary ionization model from~\cite{ramanathan_ionization_2020} at a temperature of $100~{\rm K}$. Note that free-electron approximation dramatically overestimates the cross section for low charge ionization, $Q\leq 4~e^-$.%, while for larger $Q \gtrsim 6~e^-$, the cross section is overestimated in the free-electron model by roughly 50\%. 

    \subsection{Boosted dark matter}
        \begin{figure}[!t]
            \centering
            \includegraphics[width=\linewidth]{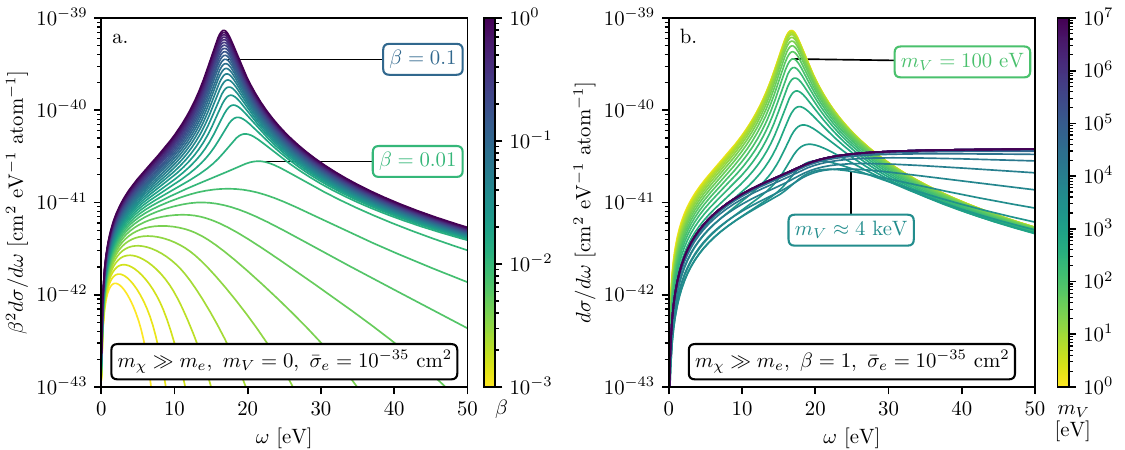}
            \caption{\textbf{Emergence of plasmon with $\beta$ and $m_V$.} Panel a shows the differential cross section per atom, weighted by $\beta^2$, for a dark matter particle with $\bar{\sigma}_e = 10^{-35}~\mathrm{cm^2}$ interacting with a silicon via a light mediator. Different colors indicate the speed of the incoming dark matter particle. The plasmon peak appears in the differential cross-section as the speed of the dark-matter particle is increased, and dominates the cross section for $\beta \gtrsim 0.01$. Panel b shows the differential cross section per atom for a boosted dark matter particle with $\beta = 1$ for various mediator masses, $m_V$. The dark matter particles interacting via lighter mediators have a cross section that is dominated by the plasmon whenever $m_V\lesssim k_F$. For $m_V\to 0$, the cross section agrees with that for millicharged-particle scattering shown in \cref{fig:mcps}. These plots are made with the Lindhard approximation for the silicon dielectric function.}
            \label{fig:boostedDM}
        \end{figure}

        \begin{figure}[!t]
            \centering
            \includegraphics[width=\linewidth]{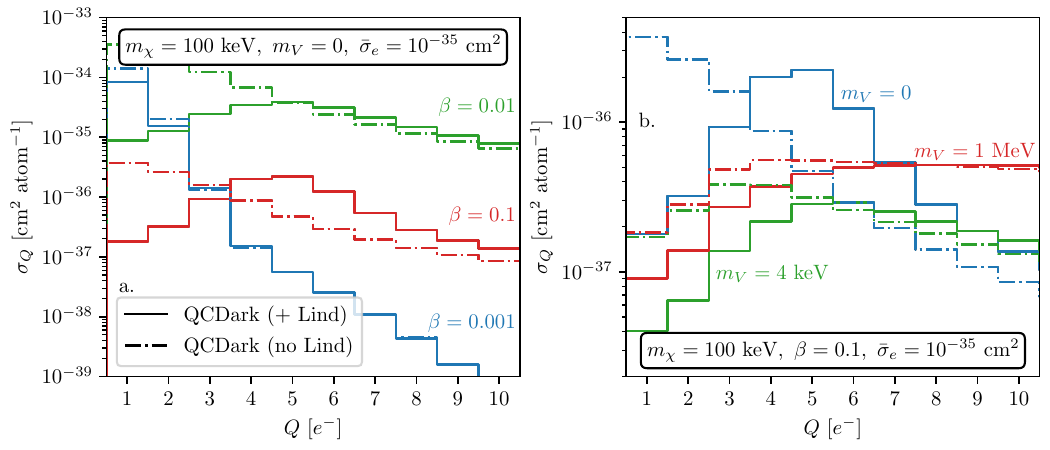}
            \caption{\textbf{Emergence of plasmon in terms of $Q$ with $\beta$ and $m_V$.} The dark matter--electron scattering cross section per atom, as a function of electron-hole pairs ionized, $Q$, following the ionization model from~\cite{ramanathan_ionization_2020} for various incoming particle speeds, $\beta$. The solid lines show the effect of including collective effects, i.e., using $\mathrm{Im}\left\{-\epsilon(\omega, k)^{-1}\right\}$, as compared to excluding them (dash--dotted lines), i.e., using only $\mathrm{Im}\left\{\epsilon(\omega, k)\right\}$. The collective effects screen the cross section at low $Q$. Panel a. shows the variation of the cross section with $\beta$; for high $\beta \gtrsim 0.01$, the plasmon becomes accessible and enhances the cross section. Panel b. shows the variation of the cross section with $m_V$, where the plasmon excitation becomes apparent for $m_V \lesssim 4$ keV. These plots are made using the imaginary part of \texttt{QCDark}, where collective effects are modeled using \cref{ansatz}.}
            \label{fig:boostedDM_ionized}
        \end{figure}

        \begin{figure}
            \centering
            \includegraphics[width=\linewidth]{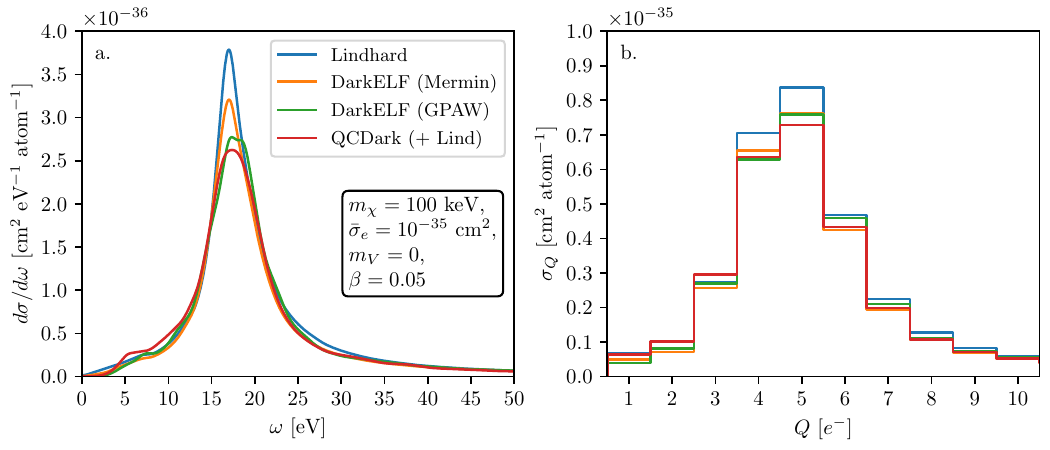}
            \caption{\textbf{Dark matter interacting with electrons via light mediator.} Panel a: the differential cross section of a 100 keV boosted dark matter particle interacting with a silicon target with $\beta = 0.05$ and a massless mediator. \textit{Blue} line corresponds to the Lindhard model~\cite{Lindhard}, \textit{orange} and \textit{green} lines corresponds to \texttt{DarkELF} (Mermin) and \texttt{DarkELF} (GPAW) respectively \cite{Knapen:2021bwg, Knapen:2021run}, and the red line corresponds to \texttt{QCDark} \cite{Dreyer:2023ovn}. Panel b: the cross section versus the number of electron-hole pairs created, where we use the secondary ionization model from~\cite{ramanathan_ionization_2020} at 100~K.}
            \label{fig:boostedDM_lin}
        \end{figure}

        The assumptions of non-relativistic nature of dark matter are usually baked into the rate equations for dark matter--electron scattering in a lattice, especially because of the low speed of dark-matter particles in the galactic halo, $\beta \lesssim 0.002$. Recent developments in the treatment of dark matter particles boosted via solar reflection and cosmic rays~\cite{An:2017ojc,Emken:2017hnp,Emken:2019hgy,Emken:2021lgc,An:2021qdl,Zhang:2023xwv,Emken:2024nox,Bringmann:2018cvk,Cappiello:2018hsu,Ema:2018bih,Dent:2019krz,Dent:2020syp} (see also~\cite{Wang:2019jtk, Xia:2020apm, Guo:2020oum, Liang:2021zkg, McKeen:2022poo}) have led to a need for a better understanding of dark matter--electron scattering without these underlying assumptions. \cref{energy-loss-V} gives the differential cross section of dark matter--electron, including the case of relativistic dark matter.

        \Cref{fig:boostedDM} shows the differential cross section of dark matter--electron scattering in a silicon target using \cref{energy-loss-V} and assuming a Lindhard model for the silicon dielectric function~\cite{Lindhard}. Panel (a) shows the differential cross section for a dark matter particle with a light mediator, $m_V \rightarrow 0$, but with varying speeds $\beta$ of the incoming dark matter particle. Note that since the minimum momentum transfer ($k_\mathrm{min}$) scales as $1/\beta$, only fast dark matter particles are able to excite the plasmon. Hence plasmons are important for $\beta \gtrsim 0.01$ and dominate the differential cross section for high-speed dark matter particles with $\beta \gtrsim 0.1$, but are largely irrelevant for halo dark matter (see also~\cite{Kurinsky:2020dpb}).

        Panel (b) of \cref{fig:boostedDM} shows the differential cross section per atom for boosted dark matter particles traveling at $\beta = 1$ interacting with electrons in a silicon lattice, but for various vector-mediator masses, $m_V$. For higher mediator masses, $m_V \gtrsim 1~{\rm MeV}$, we can approximate $k^4/\left(k^2 + m_V^2\right)^2 \sim k^4/m_V^4$, which effectively causes the high $k$ modes to be enhanced. For lighter mediator masses, $m_V \lesssim 100~{\rm eV}$, the plasmon dominates the differential cross section. 

        \Cref{fig:boostedDM_ionized} shows the dark matter--electron scattering cross section per atom as a function of ionized charge, $Q$, using the secondary 
        ionization model from~\cite{ramanathan_ionization_2020}, with and without collective effects. The collective effects \textit{`screen'} the cross section at low charge ionization $Q$, while the cross section in the high-$Q$ bins are enhanced for $\beta \gtrsim 0.01$ as the plasmon becomes accessible. Similarly, the plasmon is accessible for lighter mediators, $m_V \lesssim 4$ keV. We also note that the energy deposition peaks not at low $Q$ as it would for halo-dark matter scatters, but at $Q\gtrsim 5~e^-$; this means that the larger backgrounds towards lower values of $Q$ typically seen in current dark-matter detectors do not much impact the sensitivity of such detectors to solar-reflected and otherwise boosted dark matter. 

        \Cref{fig:boostedDM_lin} compares dark matter--electron scattering cross section per atom calculated using various dielectric functions. Panel a (b) shows the differential cross section (binned with respect to ionized charge $Q$) for a 100 keV dark matter particle of velocity $\beta = 0.05$ interacting via a massless mediator, with $\bar{\sigma}_e = 10^{-35}~\mathrm{cm}^2$. Note that the Lindhard model overestimates the cross section for the $Q = 5~e^-$ bin and there is $\sim 5-10\%$ uncertainty in the $3~e^-\leq Q\leq 7~e^-$ bins. 

        During the final stages of completing this paper, Ref.~\cite{Liang:2024xcx} appeared, which also considered the effects of the plasmon for relativistic dark-matter particles. They use the DarkELF-GPAW dielectric function and applied their formalism to cosmic-ray boosted dark matter to derive constraints using data from SENSEI at SNOLAB~\cite{SENSEI:2023zdf}.

    \subsection{Systematic uncertainties on signal predictions \label{Systematics}}

        In a counting experiment searching for hypothetical particles, the systematic uncertainties that must be understood are: 1) the uncertainty on the flux of feebly interacting particles, 2) the uncertainty on the cross section differential in energy transfer to the target, and 3) the uncertainty on the branching ratio of visible energy. The uncertainty on the flux is model dependent, and factorizes from the cross section and so we do not discuss it further. 

        As can be seen from \cref{fig:mcps}, \cref{fig:neutrinos}, and \cref{fig:boostedDM_lin} the systematic uncertainty on the cross section is modest i.e., no larger than $20\%$. When comparing the curves in \cref{fig:mcps}, \cref{fig:neutrinos}, and \cref{fig:boostedDM_lin} it is important to emphasize that we expect the Lindhard model to provide a good description of the lineshape, but not necessarily the normalization, in the vicinity of the plasmon. Since the \texttt{QCDark} curves use the Lindhard model to apply a screening correction to $\epsilon_2(\omega, k)$ we expect their current implementation to {\it underpredict} the cross section. It is therefore reasonable to use the difference between the `Mermin' and `GPAW' \texttt{DarkELF} curves as a proxy for the systematic uncertainty. In the signal bins relevant for the SENSEI millicharged-particle analysis~\cite{SENSEI:2023gie}, the uncertainty is $\lesssim 5\%$ in the relevant energy bins. 

        The cross section differential with respect to energy transfer gives an {\it upper bound} on the amount of energy deposited in the form of ionization. Energy transferred into phonons, or other vibrational modes, will likely transfer a substantial amount of their energy in the form of heat. We use the model from Ref.~\cite{ramanathan_ionization_2020} in this work, which employs a Monte Carlo method to estimate the exclusive final states. Further work characterizing the exclusive final states as a function of the deposited energy and momentum transfer would help solidify the connection between $\dd \sigma /\dd \omega$ and experimentally observable quantities such as the number of electron-hole pairs ionized. While we are currently unable to quantify this uncertainty, we do not expect any substantial changes to our qualitative findings: the plasmon peak provides helpful kinematic separation between signal and background, and serves to enhance the sensitivity of low-threshold detectors.

\section{Conclusions \label{Conclusions}}

    Semiconductors benefit from a small band gap, which allows sensitivity to small $\sim {\rm eV}$ energy depositions. Since the ionized electron-hole pairs are not free, collective effects play an important role at low energy and low momentum transfers. Of particular importance can be the role played by the bulk plasmon, a collective resonant mode that is well known to dominate EELS spectra.

    When restricted to the non-relativistic limit, many of the collective effects become less pronounced. Scattering is dominated by regions of large momentum transfer and calculations largely reduce to the evaluation of a crystal form factor~\cite{Essig:2015cda,QEDark, Dreyer:2023ovn,QCDark,Griffin:2021exdm,Trickle:2022exdm}. Relativistic and quasi-relativistic particles are crucially different as emphasized in~\cite{Kurinsky:2020dpb}. More recently, a plasmon-induced threshold effect in anisotropic ``heavy electron'' materials has recently been proposed for directional detection of light dark matter~\cite{Boyd:2022tcn}.

    In this paper, we have focused on the scattering in silicon of particles that have larger velocities than those found for virialized dark matter particles in the Milky-Way halo, including relativistic particles with $\beta \approx 1$. This allows the plasmon to be kinematically accessible at almost all energies. Since the plasmon is a longitudinal excitation, it cannot be excited by the absorption of on-shell (and therefore transverse) photons. Optical data is then a poor proxy for the response of the material to incident relativistic particles. Moreover, the many tools that have been developed for non-relativistic scattering of dark matter perform poorly in the relativistic limit, in large part due to mismodelling of the plasmon. 

    We have made use of the proper, fully relativistic formalism for energy loss in materials using realistic models/calculations for the dielectric function of silicon (our formalism can be easily applied to other materials for which the dielectric function is known). We have validated our model calculation against publicly available EELS data and find good agreement. We have applied this formalism to three characteristic models: a millicharged particle, a neutrino dipole moment, and boosted dark matter with a light mediator. In particular, the differential cross sections for the scattering in materials of millicharged particles, of neutrinos with a magnetic dipole moment, and of boosted dark matter with a dark photon mediator are given in \cref{energy-loss-fermi}, \cref{energy-loss-nu}, and \cref{energy-loss-V}, respectively. We find that the GPAW dielectric function in \texttt{DarkELF}~\cite{Knapen:2021bwg, Knapen:2021run} produces reliable results when the incoming velocity of the probe particle is high, $\beta \gtrsim 0.01$. This is because even though the dielectric function does not include all--electron effects~\cite{Griffin:2021exdm, Dreyer:2023ovn}, the target material response is dominated by low momentum transfers, $k \lesssim k_F$. 

    We have found important differences with existing implementations in the literature for the scattering of these particles. In particular, the plasmon dramatically impacts the shape of the $\dd \sigma/\dd \omega$ for all three models, and can have important consequences for how experimental searches should be optimized. The plasmon effectively acts to screen the cross section at low energy transfers (below the plasmon energy) in analogy with a Thomas-Fermi form factor in atomic physics that screens the Rutherford cross section at low energy transfer. Moreover, it enhances the differential cross section near the plasmon energy. Our results show that the proper inclusion of collective effects typically enhances a detector's sensitivity to these particles, since detector backgrounds, such as dark counts, peak at lower energies. 

\section*{Acknowledgments}

    We thank Roni Harnik and Zhen Liu for early discussions, Santiago Perez for discussions on the photo-ionization absorption model, and the broader SENSEI group for their collaboration on the NuMI search for millicharged particles. We thank Hailin Xu for catching a bug in the calculations. R.E.~acknowledges support from the US Department of Energy under Grant DE-SC0009854, from the Heising-Simons Foundation under Grant No.~79921, from the Simons Foundation under the Simons Investigator in Physics Award~623940, from the Binational Science Foundation under Grant No.\ 2020220, from Stony Brook IACS Seed Grant, and from Fermilab subcontract 664693 for the DoE DMNI award for Oscura. R.P.~is supported by the Neutrino Theory Network under Award Number DEAC02-07CH11359, the U.S. Department of Energy, Office of Science, Office of High Energy Physics under Award Number DE-SC0011632, and by the Walter Burke Institute for Theoretical Physics. R.P.~gratefully acknowledges support from the Simons Center for Geometry and Physics, Stony Brook University during the workshop {\it Lighting New Lampposts for Dark Matter and Beyond the Standard Model} where this project was initiated and a large portion of the research for this paper was performed. A.S.~is supported in part by a Stony Brook IACS Seed Grant, from Fermilab subcontract 664693 for the DoE DMNI award for Oscura, from DoE Grant DE-SC0009854, and from the Simons Investigator in Physics Award 623940. We also thank Stony Brook Research Computing and Cyberinfrastructure, and the Institute for Advanced Computational Science at Stony Brook University for access to the high-performance SeaWulf computing system, which was made possible by a National Science Foundation grant No.~1531492.

%\textbf{Author Contributions: } R.E., R.P.~and A.S.~conceived of the idea for this project together, and contributed equally to the scientific steering, writing, and conceptual elements of the project. R.P.~worked through the theory for the various cross sections considered in the `Methods' section. A.S.~performed all of the numerical calculations, including testing different publicly available codes, and comparing with experimental EELS data.  \\
%\textbf{Competing Interests: } All authors declare no competing interests whatsoever. \\
%\textbf{Data Availability: } The article uses publicly available data from \texttt{DarkELF} \cite{Knapen:2021bwg,Knapen:2021run} and \texttt{QCDark} \cite{Dreyer:2023ovn}. The EELS data is available through Refs. \cite{Baston:EELS,Ewels:2016EELSDB}, while the long wavelength electron loss function data is from Ref. \cite{Palik_book}. All results of this article are made publicly available as a supplementary data file, `Supplementary Data 1.xlsx', through the online version of the article. 
\bibliography{biblio.bib}

\appendix
\section{Additional figures for momentum-weighted energy loss function}

\begin{figure}
        \centering
        \includegraphics[width=\linewidth]{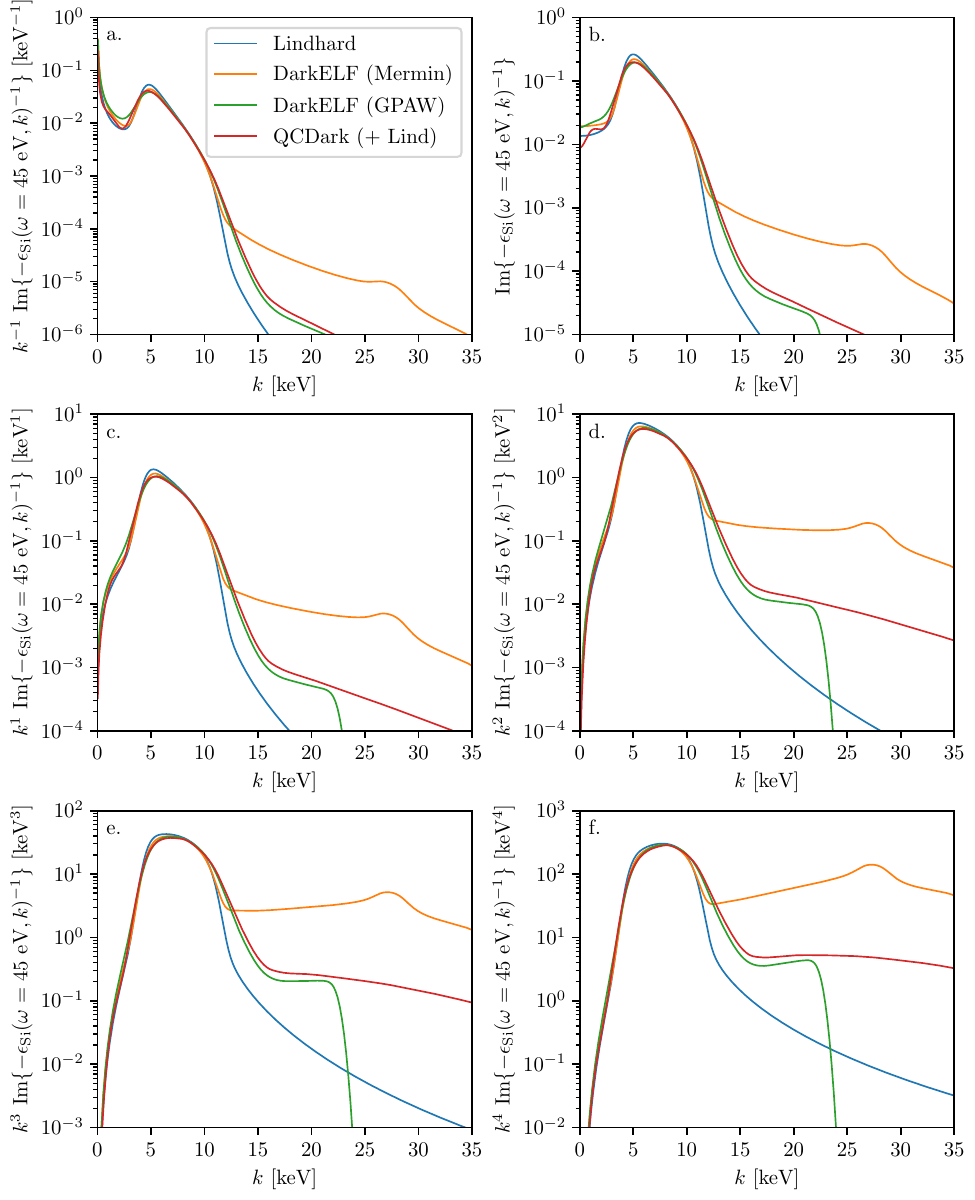}
        \caption{SupplThe wavenumber dependence of the silicon dielectric function calculated with various approximations and numerical techniques. The electron loss function, $\mathrm{Im}\{-\epsilon(\omega, k)^{-1}\}$ has been smoothed by averaging over the energy axis in a $2~{\rm eV}$ bin centered at $\omega = 45~{\rm eV}$, and underwent Gaussian smoothing on the $k$ axis with $\sigma_k = 0.5~{\rm keV}$. \texttt{QCDark} shows the effect of \textit{all-electron} inclusions at high $k \gtrsim 23 ~{\rm keV}$, while the Mermin function in \texttt{DarkELF} overestimates the imaginary part of $\epsilon(\omega, k)$ for $k\gtrsim 11 ~{\rm keV}$. }
        \label{fig:k_W}
    \end{figure}

\begin{figure}
        \centering
        \includegraphics[width=\linewidth]{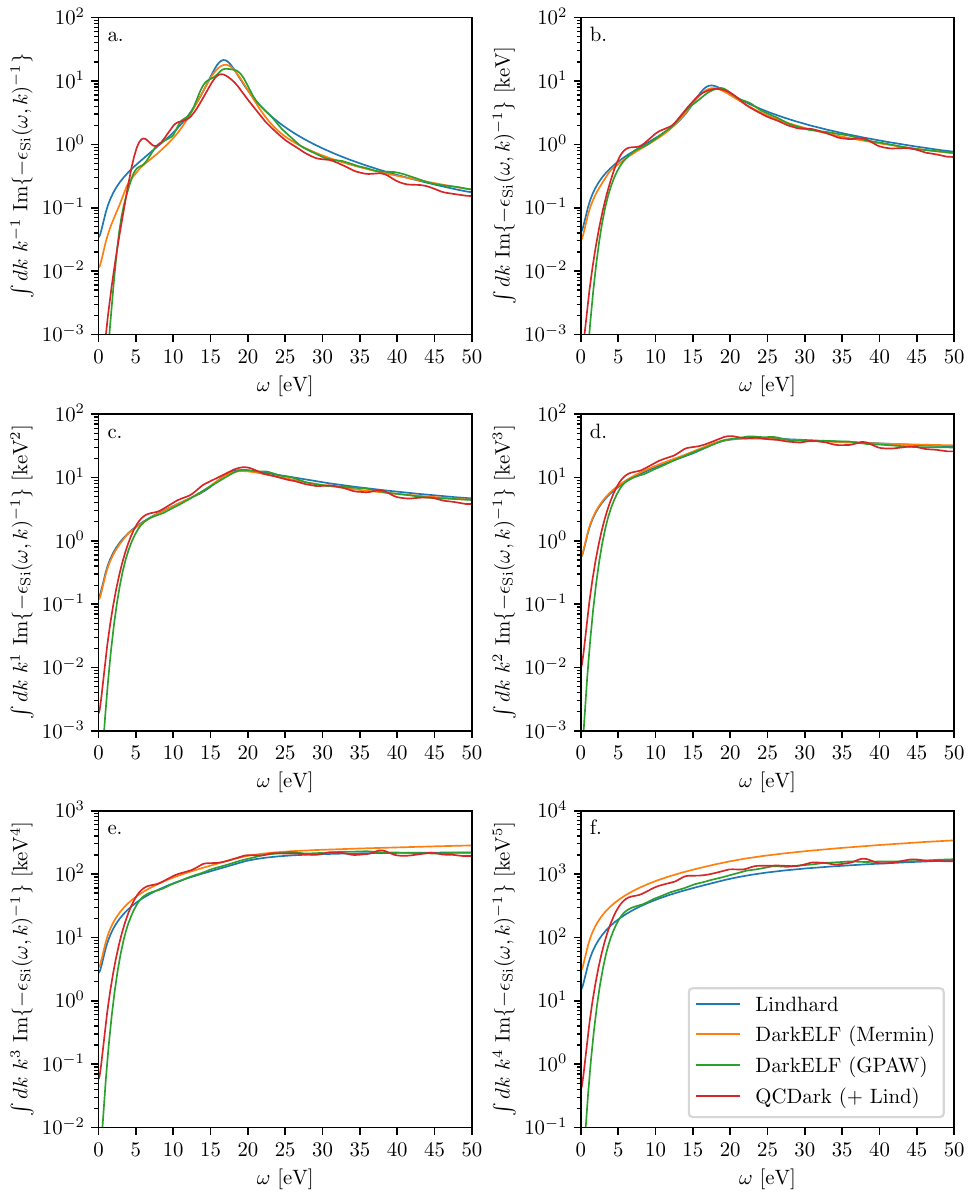}
        \caption{The frequency dependence of $\int dk\ k^n\ \text{Im}\left\{-1/\epsilon(\omega, k)\right\}$ for the silicon dielectric function calculated using various codes. The results have undergone a Gaussian smoothing with $\sigma_\omega = 0.5 {\rm eV}$. }
        \label{fig:om_W}
    \end{figure}

    In this appendix, we provide additional figures, which we feel may be helpful in interpreting our results. In \cref{fig:k_W}, we show a slice of the momentum-transfer-weighted energy loss function at a fixed energy transfer of $\omega=45~{\rm eV}$. This is the integrand in Eqs. (1) to (3) for different interactions, ranging from a massless mediator ($k^{-1}$), to a neutrino dipole moment ($k^1$), to a contact operator ($k^3$). In \cref{fig:om_W}, we show the same momentum weightings, but at the level of the integral as a function of $\omega$. These curves dominate the relativistic energy loss formulae, Eqs. (1) to (3), although they receive additional corrections from the transverse modes for relativistic kinematics. 

\end{document}